\documentclass{elsart}[12pt]
\usepackage[dvips]{graphicx}
\begin{document}
\begin{frontmatter} 
\title
{Estimate of average freeze-out volume in multifragmentation events}

\small
\author[IPNO,FLO]{S. Piantelli\thanksref{cores}},
\author[IPNO]{N. Le Neindre},
\author[IPNO]{E. Bonnet},
\author[IPNO]{B. Borderie},
\author[IPNO,INFN]{G. Lanzalone},
\author[NIPNE]{M. Parlog},
\author[IPNO]{M. F. Rivet},
\author[LPC]{R. Bougault},
\author[GANIL]{A. Chbihi},
\author[CEA]{R. Dayras},
\author[LPC]{D. Durand},
\author[GANIL]{J. D. Frankland},
\author[IPNO,ARTS]{E. Galichet},
\author[IPNL]{D. Guinet},
\author[IPNL]{P. Lautesse},
\author[LPC]{O. Lopez},
\author[NAP]{E. Rosato},
\author[LPC]{B. Tamain},
\author[LPC]{E. Vient},
\author[NAP]{M. Vigilante},
\author[CEA]{C. Volant} and
\author[GANIL]{J. P. Wieleczko},
\\
INDRA Collaboration

\address[IPNO]{Institut de Physique Nucl\'eaire, IN2P3-CNRS, F-91406
Orsay Cedex, France}
\address[FLO]{Dip. di Fisica e Sezione INFN, Universit\`a di Firenze,
I-50019 Sesto Fiorentino (Fi), Italy}
\address[INFN]{INFN, Laboratori Nazionali del Sud and Dipartimento di Fisica
e Astronomia, via S. Sofia 44, I-95123 Catania, Italy}
\address[NIPNE]{National Institute for Physics and Nuclear Engineering,
RO-76900 Bucharest-Magurele, Romania}
\address[LPC]{LPC Caen, IN2P3-CNRS/ENSICAEN et Universit\'e, F-14050
Caen Cedex, France}
\address[GANIL]{GANIL, CEA et IN2P3-CNRS, B.P. 55027, F-14076 Caen Cedex 5, France}
\address[CEA]{DAPNIA/SPhN, CEA/Saclay, F-91191 Gif sur Yvette, France}
\address[ARTS]{Conservatoire National des Arts et M\'etiers , F-75141
Paris Cedex 03, France}
\address[IPNL]{Institut de Physique Nucl\'eaire, IN2P3-CNRS et Universit\'e,
F-69622 Villeurbanne Cedex, France}
\address[NAP]{Dipartimento di Scienze Fisiche e Sezione INFN, Universit\`a
di Napoli ``Federico II'', I-80126 Napoli, Italy}
\thanks[cores]{Corresponding author: piantell@fi.infn.it} 
\normalsize
\begin{abstract}
An estimate of the average freeze-out volume for multifragmentation
events is presented.
Values of volumes are obtained by means
of a simulation using the experimental charged product partitions measured
by the
$4\pi$ multidetector INDRA for $^{129}$Xe
central collisions on $^{nat}$Sn at 32 AMeV incident energy.
The input parameters of the simulation are tuned
by means of the comparison between the experimental and simulated velocity
 (or energy) spectra of particles and fragments.
\end{abstract} 
\end{frontmatter}

A better knowledge of multifragmentation properties is of the highest importance
in the investigation of the liquid-gas phase transition in hot
nuclei~\cite{GRO97,RIC01,DAS01,BORD}. In particular, 
in various statistical and thermodynamical approaches, 
the concept of freeze-out
volume is introduced, which can be defined as the volume occupied by the
ejectiles of the multifragmenting source when their mutual nuclear
interactions become negligible.
Such a volume appears as a key quantity~\cite{BORD} and 
its knowledge is particularly important 
in the extraction of fundamental
observables such as the microcanonical heat capacity and its negative
branch or the shape of caloric curves under external
constraints~\cite{GRO97,DAG00,DAG02,CHO00}.

Up to now volume or density information at freeze-out was derived in various
ways. For example by
comparing average static and  kinetic properties of fragment distributions
with statistical multifragmentation models in which the freeze-out
volume is an input
parameter~\cite{BOND,DAG96,DES98,RA02} or from nuclear caloric curves using an
expanding Fermi gas hypothesis to extract average nuclear densities~\cite{NA02}.
In this work we present a more direct approach to determine freeze-out volumes.
Indeed dynamical simulations show that a geometrical picture is fully relevant
on the event by event basis and can be used to estimate average volumes of a
 given class of events~\cite{PA05}.
In that context
we obtained values
of the average freeze-out volume 
in multifragmentation events, from a ``fused system'' produced in central
collisions, by employing a simulation directly built event by event
from the data collected with
INDRA~\cite{POU}. At the present stage we do not want to have a fully
consistent understanding of parameter values derived from simulations
but rather a very good reproduction of data using reasonable physical
hypotheses.
Further details of the experimental and calibration
procedures may be found in Refs.~\cite{HU03,TA03}.

\begin{figure}[htbp]
\centerline{\includegraphics*[scale=0.8]
{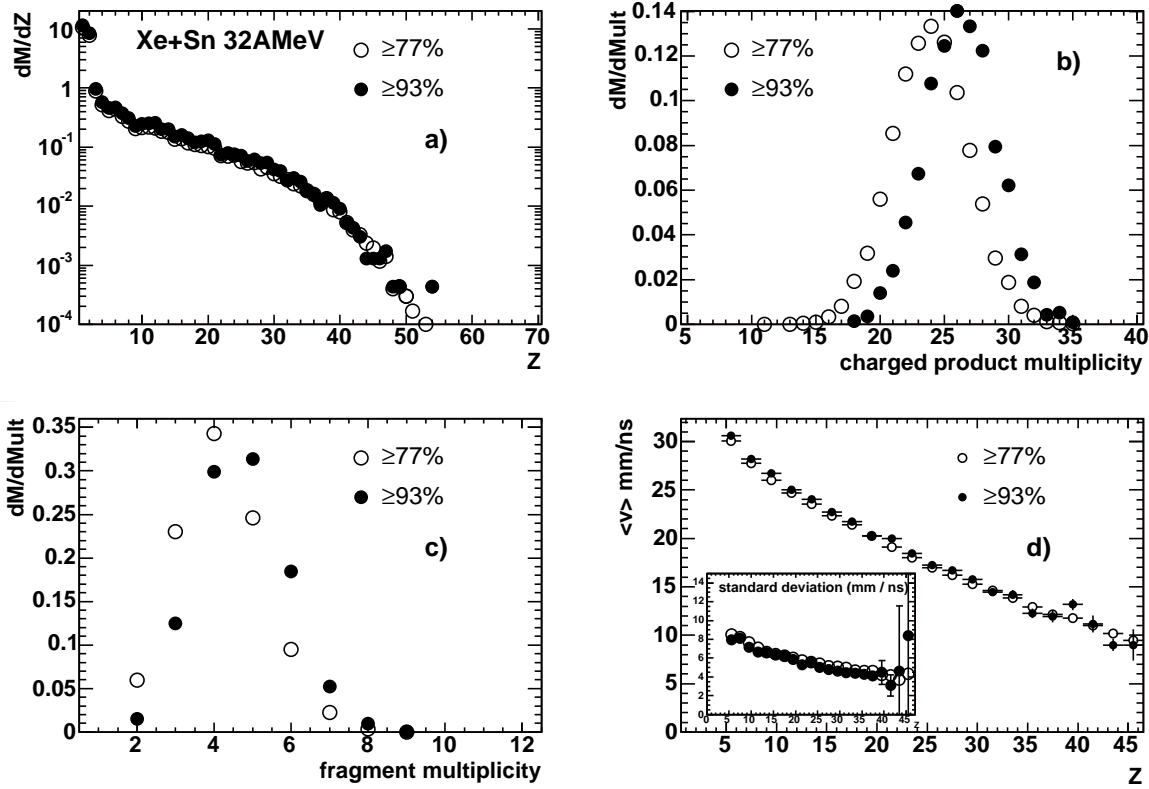}}
\caption{Properties of selected collisions: (a) differential charge multiplicity
distributions, (b-c) total charged product and fragment multiplicities and
(d) average and width (standard deviation in the inset) of the centre of 
mass velocity
spectrum of fragments as a function of their charge, regrouped by two
charge units. Open (full) symbols correspond to total detected charge greater
than or equal to 77\% (93\%) of the
entrance channel. Vertical bars are statistical errors.}
\label{Fig1}
\end{figure}

Complete experimental events (total detected charge $\geq 93\%$ and
total measured momentum $\geq 80\%$ of the entrance channel values) with
flow angle $\geq 60^{\circ}$ (corresponding to compact single source
reactions \cite{LE94,MA97,FR1}) for the reaction $^{129}$Xe+$^{nat}$Sn at
32 AMeV were
selected. The requested completeness on the total detected charge, more
severe than
that usually employed by the INDRA collaboration
($\sim 80\%$)~\cite{HU03,TA03}, was justified by
the necessity of a freeze-out source as close as possible to the
reality as input for the simulation, to avoid underestimations of the total
Coulomb repulsion among fragments and particles.
Main properties of selected multifragmenting sources are summarized in 
figure \ref{Fig1}. Depending on the required completeness on the total
detected charge ($\geq 93\%$ and $\geq 77\%$), average charged product
 multiplicity varies from 23.8 to 26.2 whereas fragment multiplicity
(with charge $\geq 5$) increases from 4.13 to 4.72.
One can also verify that the
largest completeness does
not introduce substantial bias on relevant observables 
as the differential charge multiplicity distribution, the average
experimental velocity of fragments
or the width of their velocity spectra. Moreover
flow angle distributions for both selections (not shown) exactly superimpose.

The multifragmenting source is reconstructed in the reaction centre of mass
from all the fragments ($Z\geq 5$)
and twice the particles ($Z=1,2$) and light fragments
($Z=3,4$) emitted in the range $60^{\circ}$-$120^{\circ}$ in order
to partially exclude 
pre-equilibrium emission~\cite{FR2}. Fast isotropic emission can not
be removed from the source without any theoretical assumption and, in that
respect, calculated excitation energy, mass and
freeze-out volume of the source
should be considered as upper limits.
Atomic mass of detected fragments
($Z\geq 5$) was calculated from the EAL formula~\cite{CHAR}.
The number of neutrons (which are
undetected) was calculated to keep the N/Z ratio of the entrance 
channel~\cite{HU03}.
With such a procedure the average atomic mass and atomic number of the
reconstructed sources are respectively
217 and 91. Then, the partition at freeze-out for each event was built
by ``dressing''
detected fragments proportionally to their measured charge
(calculated mass for neutron ``dressing'')
with certain percentages of detected (calculated) particles, light fragments
and neutrons. Those percentages constitute one parameter of the simulation.
The dressed fragments, assumed to be spherical, and the remaining
particles and light fragments, if any, were placed in a compact
configuration with a minimum distance $D_{min}$ among the surfaces. It was
realized by first putting particles and fragments at random on the surface of
a big sphere with volume equal to
30$V_0$ ($V_0$ being the volume of the source at normal density) and then
moving fragments and particles one by one toward the
centre of the sphere by homothetic steps. $D_{min}$ is
another parameter of the simulation. The radius of each fragment was
calculated according to the formula \(R=r_0A^{\frac{1}{3}}\), where A is the
fragment mass and $r_0$=1.2 fm; for light particles (light fragments)
experimental radii (deduced $r_0$) summarized in table \ref{table1}
were taken.
The radius for neutrons was chosen equal to the proton radius.
\begin{table}
\caption{Radii for light particles (H, He) and $r_0$ for light fragments,
from \cite{TAB}}
\begin{center} 
\begin{tabular}{cccccccccc}
\hline proton & deuteron & triton &$^3He$&$\alpha$&$^6He$&Li&Be&B&C\\ 
\hline 
1.03fm&2.8fm&2.2fm&2.4fm&2.2fm&2.4fm&1.7fm&1.5fm&1.45fm&1.3fm\\
\hline 
\end{tabular} 
\end{center}
\label{table1}
\end{table}

For each event a consistent calorimetry was made to derive the excitation energy
$E^{*}_{s}$ of the corresponding source which undergoes multifragmentation.
Thus the partition between internal excitation energy (for fragments) and
kinetic energy at freeze-out was determined.
To do that a variation of the
level density for fragments was introduced. The
level density is expected to vanish at high excitation energies~\cite{DEAN}
and the formalism adopted here is that proposed in~\cite{KOO}, where the
level density at excitation energy $\epsilon$ is expressed as
the Fermi gas level density $\rho^{FG}$ modified by a modulation factor:
{\large\begin{equation}
\rho(\epsilon)=\rho^{FG}(\epsilon)e^{-\frac{\epsilon}{T_{lim}}}
\end{equation}}

This corresponds to introducing an intrinsic temperature for fragments
$T_{frag}$ which verifies
\begin{equation}
\frac{1}{T_{frag}}=\frac{3}{2<K^{fo}>}+\frac{1}{T_{lim}}
\end{equation}
where $<K^{fo}>$ is the average kinetic energy of
fragments and particles at freeze-out and $T_{lim}$ the maximum temperature
attainable by fragments. $T_{lim}$ is a parameter of the simulation.

Equations used for calorimetry and to derive the sharing between
internal excitation energy and kinetic energy on the event by event basis
are the following:

\begin{equation}
E^{*}_{s}+\Delta B_s = \sum_{k=1}^{M_{cp}}K^k_{cp} + \Delta B_{cp}
+ M_n^{fo}<K^{fo}> + M_n^{evap}\theta_{frag} + \Delta B_n
\end{equation}

\begin{equation}
E^{*}_{s}+\Delta B_s = (M^{fo}-1)<K^{fo}>
+\sum_{k=1}^{nfrag}a_k\theta_{frag}^2
+\Delta B_{fo} + V_{Coul}^{fo}
\end{equation}

where $\theta_{frag}$ is equivalent to the temperature $T_{frag}$ in an
ensemble average.
$\Delta B$, $s$, $K$, $cp$, $n$,
and $fo$
stand respectively for mass defect, source, kinetic energy, charged products,
neutrons and freeze-out.
The neutrons evaporated from primary fragments, $M_n^{evap}$, have an average
kinetic energy along the de-excitation chain equal to the initial
temperature of fragments $\theta_{frag}$~\cite{GO89}. An internal excitation
energy
equal to \(\sum_{k=1}^{nfrag}a_k\theta_{frag}^2\)
is associated to the fragments at freeze-out,
where \(a_k=\frac{A_k}{10}MeV^{-1}\)~\cite{SHL}, $A_k$ is the mass
of the $k^{th}$
fragment and $nfrag$ is the fragment multiplicity at freeze-out.
$M^{fo}$ is the total multiplicity at freeze-out. The total kinetic energy
 \((M^{fo}-1)<K^{fo}>\) is shared
at random between all the particles and fragments at freeze-out under
constraints of conservation laws (linear and angular momentum).
$V_{Coul}^{fo}$ is the Coulomb energy of the space configuration for
freeze-out previously determined. Nuclear interactions between fragments or
particles are neglected and we shall see later that this approximation
is reasonable.
A radial collective expansion energy $E_R$ can also be introduced in equations
(3) and (4) according to the
formula \(E_R=\sum_{k=1}^{M^{fo}}({\frac{r_k}{R_0}})^{2}A_kE_0\), where
$R_0$ is the rms of fragment
distances to centre at the freeze-out volume, $E_0$ the radial expansion
energy at $R_0$ and $r$ is the distance of the considered
particle/fragment of mass A from the centre of the fragmented source.

Particles and fragments were then propagated under the effect of their
mutual Coulomb repulsion; during propagation fragments de-excited, by means of
an algorithm largely inspired by the SIMON code~\cite{DU92,SIM}.
The main differences concern the tuning of
the emission time during the evaporation sequence in order to reproduce the
results of theoretical
calculations for neutron emission~\cite{BORD2} and the constraint that the
 evaporated
particles are inside the list of the particles placed on each fragment at
freeze-out. The used emission barriers come from experimental data
(\cite{RIV} for $Z=1,2$ and \cite{VAZ} for $Z=3,4$).
In this way at the end of the de-excitation phase we obtain 
secondary charge (and mass) distributions for fragments close to the
experimental (calculated) ones; final charges (masses) are recovered, within
two charge units (four mass units), for 98\% (85\%) of fragments.
Finally experimental angular and energy resolutions were taken into account.

The last step was the comparison between experimental and simulated
spectra both for the energy of the particles and for the velocity of the
fragments (average and width) as a function of their charge, in order
to tune the four
parameters of the simulation ($E_0$, $D_{min}$, $T_{lim}$ and the percentages of
evaporated particles). We chose to compare the velocity spectra instead of
the energy ones because the velocity is less sensitive to the final mass of
fragments at the end of the de-excitation process.
The explored range for the percentage of evaporated particles was between 0\%
and 100\% (no free particles at freeze-out); 30\% was suggested by \cite{HU03}.
For $D_{min}$ we investigated an interval ranging from 0 fm (maximum possible
approach without overlap) up to 5 fm; for $E_0$ we tested from 0 (no collective radial
energy) to 1.2 MeV. Finally the explored values for $T_{lim}$ ranged among
6 MeV and 20 MeV; in
previous studies on the same or similar sample of events \cite{FR2,LAN},
$T_{lim}$ values in the range 10-12 MeV were derived.
The limiting temperature influences mainly the width (standard deviation) of
the velocity spectra; the percentage of evaporated particles controls
all the studied observables (both the standard deviation and the average
value of the fragment velocity spectra and also the energy spectra of light
particles). The distance among the nuclei surfaces at freeze-out and the
radial collective energy control mainly the average of the fragment velocity
spectra and, more weakly, the particle energy spectra;
$D_{min}$ and $E_0$
are correlated, indeed
a larger surface
distance implies a weaker Coulomb repulsion which can be compensated for by
a larger radial kinetic energy.

A $\chi^2$ minimization procedure was used to determine the best fit to
the data. To reduce the total influence of particles and light fragments but
emphasize their high energy tails very sensitive to freeze-out emissions,
$\chi^2$ was calculated using all the fragment velocity spectra and the
Log of particle and light fragment energy spectra.
The best agreement, corresponding to a $\chi^2$=0.953, was obtained
for the following set of parameters:
$T_{lim}$=10 MeV, 90\% of evaporated particles and light fragments,
$D_{min}$=2 fm and $E_0$=0.6 MeV.
The obtained values for the percentage of evaporated particles are
 larger than those extracted in \cite{HU03}; this discrepancy may
be due to some intrinsic lack of sensitivity of the method proposed
in \cite{HU03}.
Indeed fragment-particle correlations are only fully sensitive when
fragments de-excite
with a sufficient distance between them and in that sense it is
 impossible to go back up to the freeze-out configuration. In
any case the values presented in \cite{HU03} constitute a reliable inferior
limit for the percentage of evaporated particles. 
The total average mass of dressed fragments
is 208, which corresponds to 96\% of the mass of the source at freeze-out.

Note that with the retained parameters
the calorimetry procedure gives an average excitation energy
(thermal+collective) $<E_s^*>$=6.7
AMeV for the source which undergoes multifragmentation, which leads,
for fragments, to a temperature
\mbox{$T_{frag}$=$<\theta_{frag}>$=$6.3$MeV} and an average excitation
energy of 3.9 AMeV.
The average kinetic energy of fragments and particles at freeze-out
corresponds,
if interpreted in terms of ``kinetic temperature'', to a value $T_{kin}$ of
17.5 MeV.
A deep understanding of those different  numbers and their relation with
the observed fragment partitions is out of the scope of the present letter.
However we can mention three possible explanations that one will have to
discuss in a near future. One is fully related to
the procedure followed here, assuming thermal equilibrium for fragments, which
reveals the major role of a limiting
temperature (for fragments); in that context the influence of $T_{lim}$ on
partitions in microcanonical multifragmentation model like~\cite{RA02}
will have to be checked in details.
A second one refers to a fast fragmentation for which particles and fragments are
early emitted: average primary kinetic energies are then related to the
Fermi momentum~\cite{GO74,LU03,LA04}.
Finally the ``kinetic temperature'' deduced could also reflect the fact that
few very energetic particles emitted during the
expansion-thermalization phase~\cite{FR90}
significantly increase the average kinetic energy related to $T_{kin}$.

\begin{figure}[htbp]
\centerline{\includegraphics*[scale=0.65]
{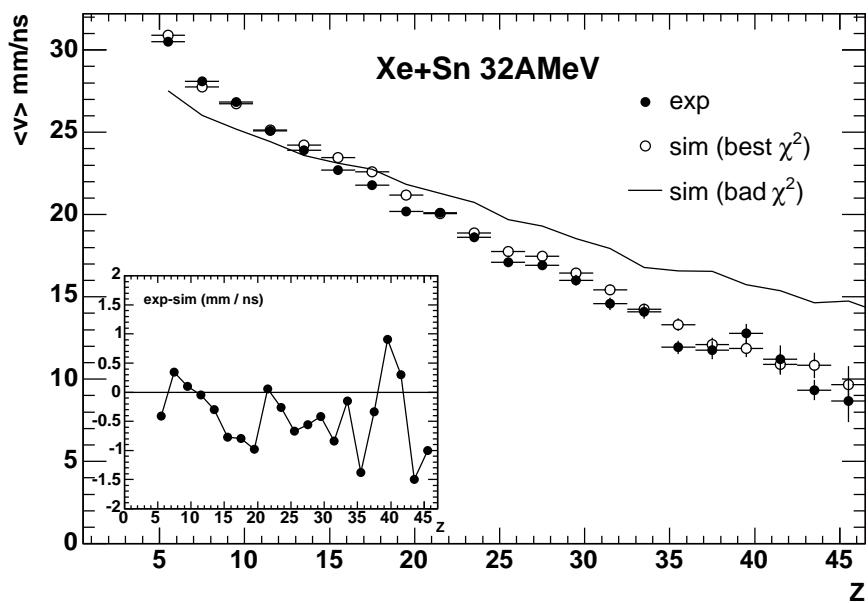}}
\caption{Experimental (full symbols), best simulated (open symbols) and
bad simulated (line, see text) centre of mass average velocity of
 fragments
as a function of the final fragment charge, regrouped by two charge units.
Vertical bars are statistical errors. The inset shows the
absolute difference between the experimental data and the best simulation.}
\label{Fig2}
\end{figure}

\begin{figure}[htbp]
\centerline{\includegraphics*[scale=0.4]
{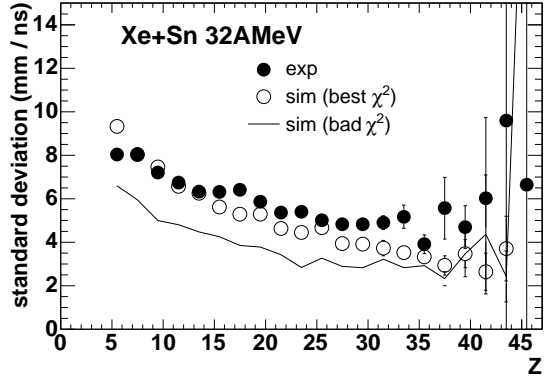}}
\caption{
Standard deviation for the 
centre of mass velocity spectra of fragments
as a function of the fragment secondary charge, regrouped by two charge units
(symbols and line as in figure \ref{Fig2}).
Error bars represent errors coming from the fits.}
\label{Fig3}
\end{figure}

\begin{figure}[htbp]
\centerline{\includegraphics*[scale=0.65]
{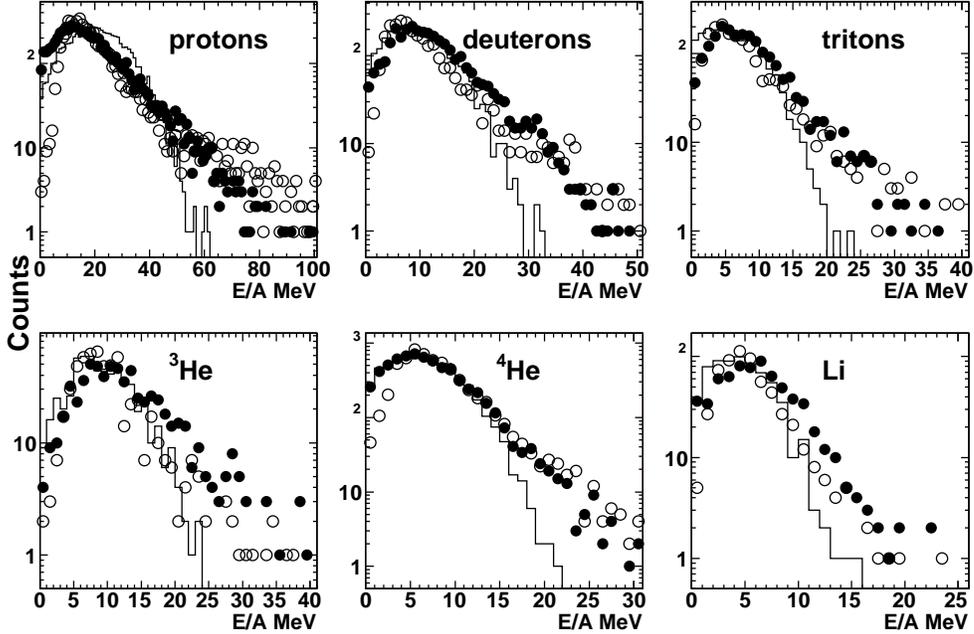}}
\caption{Center
of mass energy spectra per nucleon for light particles and lithium emitted with
polar angle in the centre of mass $60^{o}\leq \vartheta^{cm}\leq 120^{o}$
(symbols as in figure \ref{Fig2} and histograms for bad $\chi^2$ simulation).}
\label{Fig4}
\end{figure}

In figure \ref{Fig2} the experimental centre of mass average velocity of the
fragments
(full symbols) is compared to the best simulation (open symbols) as a
function of the final fragment charge. From the inset it is possible to
appreciate the small absolute gap between the experimental data 
and the simulation. Coulomb repulsion at freeze-out contributes to
$\sim$70-80\% of the calculated average velocities.
The standard deviation of the fragment centre
of mass velocity spectra
as a function of the fragment charge for the
experimental data (full symbols) and the best simulation (open symbols) is
presented in figure \ref{Fig3}. 
Spatial configurations at freeze-out and fragment decays are
only responsible for $\sim$60-70\% of the observed widths and 
the introduction of a limiting temperature in the
simulation turned to be mandatory to account for the experimental values.
In figure \ref{Fig4} the experimental and best simulated centre of mass energy
spectra for particles and lithium fragments are presented. The agreement
between
simulation and data is good.
Relative angles and velocities between fragment pairs were also compared
through relative velocity correlation functions (not shown) and exhibit a
reasonable agreement between experimental and simulated data.

Before obtaining estimates on freeze-out volume, we can discuss the
error bars 
on the different parameters of the simulation. The standard method was
used: errors are
extracted from simulations for which  $\chi^2$ = best$\chi^2$ + 1. The derived 
ranges for parameters are the following:
9-11 MeV for $T_{lim}$, 70-90\% of evaporated particles and light fragments,
0.5-3 fm for $D_{min}$ and 0.3-0.6 MeV for $E_0$.
As previously noted, only those two last parameters are strongly correlated.
Therefore, the same reasonable $\chi^2$ is obtained for couples of extreme values like
0.5 fm-0.3 MeV and 2 fm-0.6 MeV keeping constant $T_{lim}$ and the
percentage of evaporation.
Note that even in simulations with $D_{min}$=0.5 fm 
average distances among the surfaces of nuclei equal to or larger than 2.4 fm
are obtained,
compatible with the freeze-out definition as the configuration where the
nuclear force among the primary products vanishes.
Finally to appreciate the sensitivity of parameters, as an example, results
from a simulation with $\chi^2$ = best$\chi^2$ + 3 are also displayed
in figures \ref{Fig2}, \ref{Fig3} and \ref{Fig4}. For that
simulation three parameters have values close to or at the limits of the error
bars just
discussed before, $T_{lim}$=11 MeV, $D_{min}$=0 fm and $E_0$=0.34 MeV, and one
has a value largely outside: 0\% evaporation. It corresponds to a
compact freeze-out state containing all the detected charged products and
calculated number of neutrons (sphere volume of 7.4$V_0$ - see definition
below).
We note a wrong slope in the spectrum of the
average velocities of fragments versus their charge (figure \ref{Fig2}) mainly
related to the freeze-out topology imposed by the large number of particles
(the heavier the fragment, the larger its distance relative to the center of
the volume). The width of the velocity spectra are
underestimated (figure \ref{Fig3}) due to the absence of smearing effect from
evaporation. 
The energy spectra of particles and light fragments (figure \ref{Fig4})
have steeper slopes than the experimental ones because of the
absence of evaporation (no boost from primary fragments).

\begin{figure}[htbp]
\centerline{\includegraphics*[scale=0.4]
{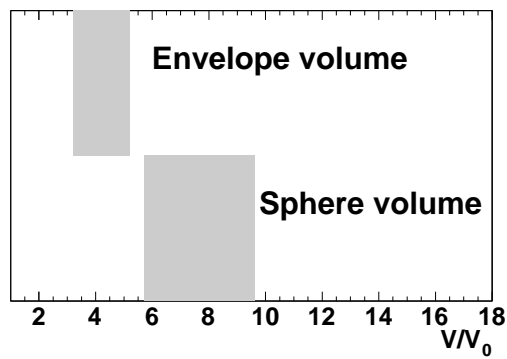}}
\caption{Estimated average freeze-out volume in units
of $V_0$. Envelope volumes refer to realistic volumes (see text) and sphere
volumes correspond to spheres including all the nuclei at freeze-out.}
\label{Fig5}
\end{figure}

Once the values of parameters and their range defined, the corresponding
freeze-out volumes were calculated from the envelopes of nuclei at freeze-out
(before starting the Coulomb propagation). The envelope was defined in such a
way that two external adjoining spherical nuclei $A_{i}$ and $A_{j}$ (whose
centres are located at distances $d_{i}$ and $d_{j}$ from the
centre of the fragmenting source)
 are linked by a portion of sphere with same centre but a
radius $R_e$ = 1/2 ($d_i$ + $d_j$ + $r_{0}A_i^{1/3}$ + $r_{0}A_j^{1/3}$).
It is worthwhile to recall that the volume is influenced by two parameters
of the simulation: the minimum
surface distance $D_{min}$ and the percentage of evaporated particles, and by
the compact configuration built at random.
The
technique used to calculate the volume of the envelope was to fill up uniformly
with points a sphere including all
the nuclei and to calculate the ratio among the number of points included
inside the envelope and the total number of points inside the sphere. In
this way it is possible to obtain a good representation of the effective
volume occupied by the nuclei also for non-spherical configurations. For the
best simulation the average envelope volume is 4.2$V_0$; the
volume of the sphere including all the nuclei is an overestimation
and, on average, for the best simulation, it is 7.65$V_0$. Volume estimates,
taking into account error bars on parameters, are presented in
figure~\ref{Fig5}. Envelope volumes range from 3.2 to 5.2$V_0$ and sphere
volumes from 5.7 to 9.6$V_0$. A small increase ($\sim$10\%) of envelope
volumes is observed  when the fragment multiplicity increases from 2 to 8 whereas
sphere volumes keep constant whatever the multiplicity.
Standard deviations
of envelope volumes generated by the simulation realized event by event 
are close to 10\% of average volumes, which qualitatively  agree with
predictions, in the coexistence region, of a microcanonical lattice gas model
with a constrained average volume~\cite{CHO00,GU04}.

To summarize, with the powerful $4\pi$ multidetector INDRA, we were able to 
obtain information on the
freeze-out volume in single source multifragmentation events 
 in the framework of a simulation using the experimental data.
The presented method needs data with a
very high degree of completeness, which is crucial for a good estimate of
Coulomb energy. 
The use of the widths of fragment velocity spectra in the comparison between
data and simulation shows that the
introduction of a limiting temperature in the range 9-11 MeV seems mandatory.
Work is in progress to study the evolution up to
50 AMeV incident energy of
freeze-out volume
for the same system and to
derive consistent and reliable information from parameters
of simulations~\cite{PI04}.


\end{document}